\documentclass[12pt]{article}
\usepackage{graphicx}
\usepackage{amsmath}
\usepackage{longtable}

\begin{document}

\makeatletter
\renewcommand*{\@cite}[2]{{#2}}
\renewcommand*{\@biblabel}[1]{#1.\hfill}
\makeatother

\title{Using the reduced proper motions of Tycho-2 stars with B-V from 0.75 to 1.25: Monte Carlo simulations}
\author{George~A.~Gontcharov\thanks{E-mail: georgegontcharov@yahoo.com}}

\maketitle

Pulkovo Astronomical Observatory, Russian Academy of Sciences, Pulkovskoe sh. 65, St. Petersburg, 196140 Russia

DOI: 10.1134/S1063773709090072 \\
Received November 6, 2008

The stellar composition of the Tycho-2 Catalogue in the range $0.75<B-V<1.25$ has been reproduced through Monte Carlo simulations. 
For young and old stars of the red giant clump (RGC), the red giant branch, subgiants, red dwarfs, and thick-disk giants, we have specified the distributions in
coordinates, velocities, $B-V$, and $M_V$ as a function of $B-V$ and calculated their reduced proper motions, photometric distances from the $(B-V)-M_V$ calibration, 
and photoastrometric distances from the reduced proper motion -- $M_V$ calibration. Our simulations have shown the following: (1) a sample of thin-disk giants
within 500 pc with an admixture of less than 10\% of other stars can be produced; (2) a sample of dwarfs within 100 pc almost without any admixture of other stars 
can be produced; (3) the Local Spiral Arm affects the RGC composition of any magnitude-limited catalog in favor of giants younger than 2 Gyr;
(4) the samples produced using reduced proper motions can be used for kinematic studies, provided that the biases of the quantities being determined are simulated 
and taken into account; (5) the photometric distances correlate with the photoastrometric ones because of the correlation between the proper motion
and magnitude; (6) the photometric distances are closer to the true ones for the red giant branch and red dwarfs as the categories of stars with a clear $(B-V)-M_V$ relation, 
while the photoastrometric distances are closer to the true ones for the RGC, subgiants, and thick-disk giants; (7) the calculated distances differ
systematically from the true ones, but they can be used to analyze the three-dimensional distribution of stars. 
Our simulations confirm the validity of our previous selection of RGC stars from Tycho-2.

\newpage

\section*{Introduction}

In recent years, multicolor photometry and proper motions from modern large-scale catalogs, such as Tycho-2 (H\o g et al. 2000), 2MASS (Skrutskie et al. 2006), and 
UCAC2 (Zacharias et al. 2004), have been used to classify stars and to calculate the interstellar extinction, distances, and spatial distribution of stars 
(Rocha-Pinto et al. 2004; Veltz et al. 2008). The greatest variety of stars is observed in the range of color indices $B-V=0.75^m\div1.25^m$, i.e., the
spectral types G and K: white dwarfs, subdwarfs, red dwarfs (RD), subgiants (SG), the red giant branch (RGB), young and old stars of the red giant clump
(RGC-Y and RGC-O, respectively), supergiants, and thick-disk and halo stars.

Figure 1a shows the positions of 16726 Hipparcos stars (ESA 1997) with the most accurate data on the $(B-V)$ -- $M_J$ diagram (in what follows, the $J$ and $K_s$
magnitudes were taken from the 2MASS Catalogue). The vertical straight lines highlight the range under consideration.

In large-scale catalogs, the color index correlates noticeably with the apparent magnitude for faint stars near the Galactic plane. This allows giants and
dwarfs to be separated because of their significantly different distances from the Sun and reddenings: at $V=14^m$, a giant with $M_V\approx1^m$ is located at a distance
of several kpc from the Sun and will redden strongly, in contrast to a dwarf with $M_V\approx6^m$ located at a distance of $\approx400$ pc. This can be seen
from Fig. 1b, which shows the positions of 35139 2MASS stars located in the region with Galactic coordinates $l\approx135^{\circ}, b\approx0^{\circ}$ on the $(B-V)$ -- $J$
diagram (we assumed that $(B-V)=1.464(J-K_s)$). The few Tycho-2 stars in 2MASS lie to the left and above the slanting line shown in the figure. As a result
of the correlation between the $J$ magnitude, distance, and reddening, the most numerous (in the sample under consideration) main-sequence (MS) A--F stars
(the points along the left curve), nearby dwarfs (along the middle curve), and RGC giants (along the right curve) can be separated confidently.

However, this approach is inapplicable for the Tycho-2 stars and, in general, for the stars of the nearest kiloparsec: Fig. 1c shows the positions of
21144 Tycho-2 stars with accurate photometry located within 20$^{\circ}$ of the Galactic poles (low reddening) on the $(B-V)$ -- $J$ diagram. The slanting line indicates
the magnitude limitation for the stars with accurate photometry. We see that nearby dwarfs are inseparable from distant giants for the indicated $B-V$ range if
only the photometry is used.

When the Tycho-2 stars are investigated, the absolute magnitude can be replaced with the reduced proper motion $M'_{V}$ if the systematic stellar motions,
observational selection, and extinction are taken into account (Parenago 1954; Jones 1972):
\begin{equation}
\label{mva}
M'_V=V+5+5\log(\mu)-A_V,
\end{equation}
where $V$ is the magnitude, $A_V$ is the extinction in the $V$ band, and $\mu=((\mu_{\alpha}\cos\delta)^2+\mu_{\delta}^2)^{1/2}$ is the proper
motion expressed in arcseconds per year.

Based on the formula
\begin{equation}
\label{rr}
\log(r)=(V-M_V+5-A_V)/5,
\end{equation}
where $r$ is the distance and $M_V$ is the absolute magnitude in the $V$ band, we can calculate the distances of stars using a statistical relation between $M_V$
and $M'_V$, for example, a linear one:
\begin{equation}
\label{mm}
M_V=aM'_V+b,
\end{equation}
where $a$ and $b$ are some coefficients. We will call these distances \emph{photoastrometric} ones (Robin et al. 2003),
in contrast to the \emph{photometric} distances calculated from Eq. (2) based on a statistical relation between the absolute magnitude and color index, for example,
\begin{equation}
\label{mbv}
M_V=c(B-V)+d,
\end{equation}
where $c$ and $d$ are some coefficients.

In recent years, the reduced proper motions have been used in many studies: by Rocha-Pinto et al. (2004) to separate distant giants and nearby M-type
dwarfs from 2MASS, by Belikov et al. (2002) to calculate the photoastrometric distances of Tycho-2 stars in the Per OB2 association, by Rybka (2007)
and Gontcharov (2008b) to select RGC stars from Tycho-2, by Gontcharov (2008a) to select O-B stars from Tycho-2 and to calculate their photoastrometric
distances, and by Veltz et al. (2008) to separate giants and dwarfs with $J-K_s=0.5^m\div0.7^m$ near the Galactic poles.

\begin{figure}
\includegraphics{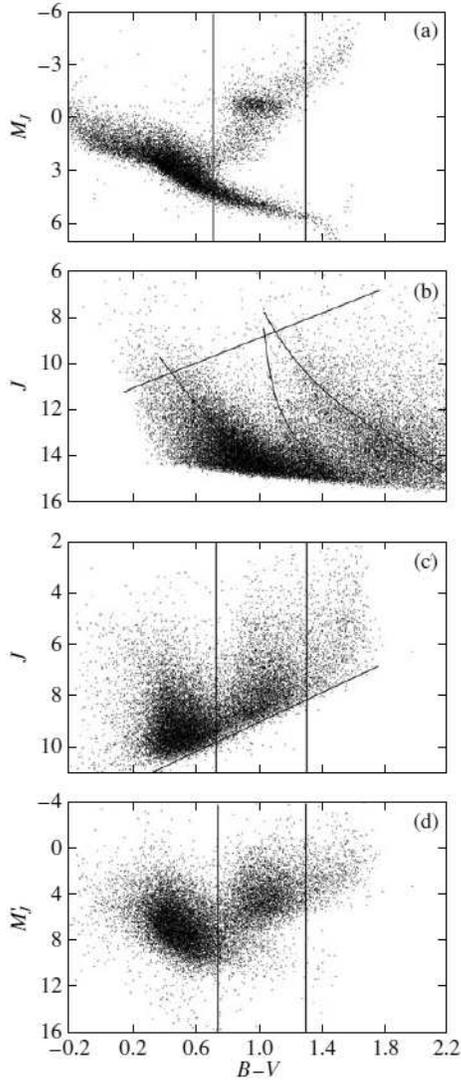}
\caption{(a) 16726 Hipparcos stars with the best data on the $(B-V)$ -- $M_J$ diagram, 
(b) 35139 2MASS stars with $l\approx135^{\circ}, b\approx0^{\circ}$ on the $(B-V)$ -- $J$ diagram,
(c) 21144 Tycho-2 stars with accurate photometry and $|b|>70^{\circ}$ on the $(B-V)$ -- $J$ diagram, 
(d) 21144 Tycho-2 stars with accurate photometry and $|b|>70^{\circ}$ on the $(B-V)$ -- $M'_{J}$ diagram. 
The vertical straight lines mark the $(B-V)$ range considered here.
}
\label{bv}
\end{figure}

Figure 1d shows the positions of 21144 previously mentioned Tycho-2 stars on the $(B-V)$ -- $M'_J$ diagram. 
At first glance, the RGC giants are separated here from the few dwarfs. However, some mutual ``contamination'' of these samples is possible.

In this paper, we make an attempt to determine, through Monte Carlo simulations, to what extent the use of the reduced proper motion implemented
in practice by Gontcharov (2008b) is justified for the classification of stars, in particular, for the separation of giants and dwarfs in the range $B-V=0.75^m\div1.25^m$, 
and to what extent the photoastrometric and photometric distances differ from the true ones.

Note that, according to Fig. 1d, Tycho-2 contains a small number of supergiants in the range $B-V=0.75^m\div1.25^m$, while the subdwarfs and white dwarfs
were mostly excluded by the magnitude limitation of this catalog. However, an appreciable number of subdwarfs with $(B-V)<0.75^m$ are seen several magnitudes
below the MS. This isolation of subdwarfs from the MS results from their belonging to a Galactic population that is barely involved in the Galactic
rotation and, hence, has a high velocity relative to the Sun in the direction opposite to the Galactic rotation (the so-called asymmetric drift), which increases noticeably $M'$
(Jones 1972). Thus, the subdwarfs will be selected from Tycho-2 comparatively easily and are a good material for a separate study.

The thick-disk (TDG) and halo giants also have a large asymmetric drift that increases $M'$. Therefore, they should be considered in this study as separate
categories. However, according to the Besan\c{c}on model of the Galaxy (Robin et al. 2003), the ratio of the number of halo giants to the number of thin-disk
stars in the solar neighborhood is about 0.0006. Simulations show that since Tycho-2 is a magnitude limited catalog, this corresponds to just 10 halo giants
for every 35000 RGC-O stars, i.e., in contrast to the catalogs with fainter stars, there are virtually no halo giants in this catalog. TDG are considered below as
a separate category with $B-V$ and $M_V$ as those for RGC-O.

\section*{The method}

Here, we consider six categories of stars: TDG, RD, SG, RGB, RGC-O, and RGC-Y. Since the RGC-Y stars are younger than 2 Gyr, their distribution 
in X can be nonuniform due to the association of these stars with the Local Spiral Arm (Gontcharov 2008b). This case is considered as an
additional seventh category of stars designated as RGC-Y*.

In our simulations, we used normal and uniform distributions realized with the Microsoft Excel 2007 random number generator, whose general description
was given by Wichman and Hill (1982). For each category, we generated 200000 model stars.

For each category, Table 1 specifies:
\begin{itemize}
\item the distributions of stars in rectangular Galactic $X$, $Y$, and $Z$ coordinates in pc (we took a uniform distribution in $Z$ for RD and a normal distribution
with the variance from Veltz et al. (2008) and Gontcharov (2008b) for the remaining categories);
\item the distributions in velocity components along the Galactic longitude $l$ and latitude $b$, $V_l$, and $V_b$, in km s$^{-1}$ from Robin et al. (2003) and Veltz
et al. (2008);
\item the distributions in $B-V$;
\item the dependence of $M_V$ on $B-V$ and the scatter of $M_V$ about this dependence.
\end{itemize}

The distribution of stars in $B-V$, the dependence of $M_V$ on $B-V$, and its scatter were taken in accordance with the database of evolutionary tracks and
isochrones by Girardi et al. (2000).

In accordance with the Tycho-2 characteristics, we took the error in the proper motion component $\mu_l$ as a function of the $V$ magnitude
\begin{equation}
\label{sml}
\sigma(\mu_l)=0.0002e^{0.3V}
\end{equation}
and the same for $\mu_b$ (arcsec yr$^{-1}$) as well as the photometric error in the $V$ band
\begin{equation}
\label{sv}
\sigma(V)=0.005e^{0.3V}
\end{equation}
and the same in the $B$ band.

Based on the quantities specified in Table 1, we calculate the following for each model star:
the true distance
$$ r=(X^2+Y^2+Z^2)^{1/2}, $$
the Galactic coordinates $l$ and $b$
$$ \tan(l)=Y/X, $$
$$\tan(b)=Z/(X^2+Y^2)^{1/2}, $$
the proper motion components
$$ \mu_l=V_l/(4.74r), $$
$$ \mu_b=V_b/(4.74r), $$
the total proper motion
$$ \mu=(\mu_l^2+\mu_b^2)^{1/2}. $$

The interstellar extinction can be determined from some model as a function of the Galactic coordinates or from multicolor stellar photometry, as in
Gontcharov (2008b). Simulating the second method and taking into account the fact that both methods yield similar results, for simplicity, we calculate the
interstellar extinction using a barometric law (Parenago 1954):
$A_V=A_0Z_{A_V}(1-e^{(-r|\sin(b)|/Z_{A_V})})/|\sin(b)|$,
where $A_0$ is the extinction in the equatorial plane per 1 pc, $Z_{A_V}$ is the characteristic half-width of the absorbing layer, and calculate the extinction error as
a quadratic sum of the photometric errors in the $B$ and $V$ bands:
\begin{equation}
\label{ext}
\sigma(A_V)=0.007e^{0.3V}.
\end{equation}
Since the stars under consideration are mostly within 1 kpc, we may take $A_0=0.0015^m$ per 1 pc and $Z_{A_V}=100$ pc. Stars with both high and negative \emph{measured}
extinctions appear because of the photometric errors.

Next, we calculate the magnitude
$$ V=M_V-5+5\log(r)+A_V. $$
Since there are almost no stars with $V>10.6^m$ and accurate photometry in Tycho-2, the model stars fainter than this magnitude are rejected.

For the remaining stars, we calculate the photometric distance $r_{ph}$ using Eq. (2), where $M_{V}$ is the function of $(B-V)$ specified in Table 1, and then the coordinates
$$ X_{ph}=r_{ph}\cdot\cos(l)\cos(b), $$
$$ Y_{ph}=r_{ph}\sin(l)\cos(b), $$
$$ Z_{ph}=r_{ph}\sin(b). $$
Subsequently, we calculate the reduced proper motion $M'_V$ from Eq. (1), the coefficients of dependence (3) by the least-squares method, the photoastrometric
distance $r_{rpm}$ from Eq. (2) by taking these coefficients into account, and, finally, the $X_{rpm}$, $Y_{rpm}$ and $Z_{rpm}$ coordinates.


\begin{table*}
\def\baselinestretch{1}\normalsize\scriptsize
\caption[]{Distributions of the categories of stars under consideration in $X$, $Y$, $Z$, $V_l$, $V_b$, $B-V$ and the dependence of $M_V$ on $B-V$.
N($a$; $b$) denotes a normal distribution with a mean $a$ and a standard deviation $b$, U($a$; $b$) denotes a uniform distribution from $a$ to $b$.}
\label{init}
\[
\begin{tabular}{llllllll}
\hline
\noalign{\smallskip}
Category & $X$, pc & $Y$, pc & $Z$, pc & $V_l$, km/s & $V_b$, km/s & $(B-V)$ & $M_{V}=f(B-V)$ \\
\hline
\noalign{\smallskip}
RGC-Y       & U(-1750;1750) & U(-1750;1750) & N(0;180)    & N(0;15) & N(0;10) & N(U(0.8;1);0.01)     & N(0.4;0.5)            \\
RGC-Y*    & N(0;500)      & U(-1750;1750) & N(0;180)    & N(0;15) & N(0;10) & N(U(0.8;1);0.01)     & N(0.4;0.5)            \\
RGC-O      & U(-1500;1500) & U(-1500;1500) & N(0;250)    & N(0;30) & N(0;15) & N(U(0.83;1.13);0.03) & N(1.67(B-V)-0.93;0.3) \\
RGB         & U(-2000;2000) & U(-2000;2000) & N(0;270)    & N(0;32) & N(0;16) & U(0.8;1.25)          & N(-5.8(B-V)+7.5;0.6)  \\
SG         & U(-750;750)   & U(-750;750)   & N(0;300)    & N(0;34) & N(0;17) & U(0.75;1.05)         & N(U(2;4);0.1)         \\
RD          & U(-200;200)   & U(-200;200)   & U(-200;200) & N(0;35) & N(0;18) & U(0.75;1.25)         & N(5.0(B-V)+1.5;0.4)   \\
TDG         & U(-1500;1500) & U(-1500;1500) & N(0;1000)   & N(0;60) & N(0;42) & N(U(0.83;1.13);0.03) & N(1.67(B-V)-0.93;0.3) \\
\hline
\end{tabular}
\]
\end{table*}


\section*{General classification of stars}

Figure 2a shows the distributions of the seven categories of model stars under consideration in $M_V$. The short dashes represent the distribution of RGC-Y* stars. 
In fact, all model stars are selected in the same region of space, but the selection due to the limitation $V<10.6^m$ for Tycho-2 stars gives completely
different numbers of stars from different categories (without allowance for the mass function to be taken into account below). When the distribution in Fig. 2a
is compared with the $M_V$ ranges specified in Table 1, the influence of the same selection in favor of higher luminosity stars is noticeable.

\begin{figure}
\includegraphics{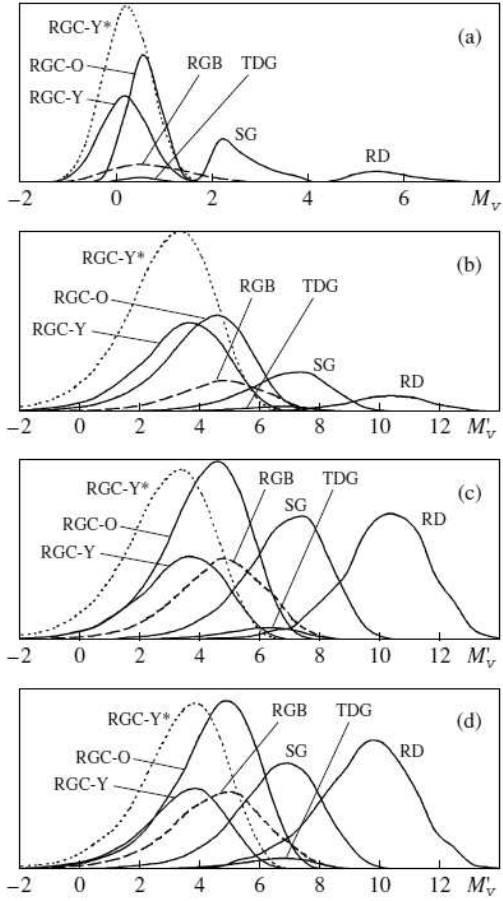}
\caption{Distributions of stars from the categories under consideration: (a) in $M_V$ , (b) in $M'_V$, (c) the same with the mass function included, and 
(d) the same with the mass function and all errors included.
}
\label{mv}
\end{figure}

For TDG, the ratio of their number to the number of thin-disk stars in the solar neighborhood is important. This ratio varies from 0.07 in the Besan\c{c}on
model of the Galaxy (Robin et al. 2003) to 0.09 in Veltz et al. (2008) and 0.15 in Soubiran et al. (2003). In this study, we took the ratio to be 0.1; given this
value, as shown in Fig. 2, the number of TDG stars is insignificant compared to the remaining categories.

Figure 2b shows the distributions of the same model stars in $M'_V$, while Fig. 2c shows the same distributions as those in Fig. 2b, but the Salpeter (1955)
mass function was taken into account: $M^{-2.35}$. The masses for the categories of stars under consideration were taken from Girardi et al. (2000): 
2 M$_{\odot}$ for RGC-Y, 1.5 M$_{\odot}$ for RGC-O, 1.3 M$_{\odot}$ for RGB, 1.2 M$_{\odot}$ for SG, 0.8 M$_{\odot}$ for RD, and 1.4 M$_{\odot}$ for 
TDG (as a mixture of RGC-O and RGB).

As we see from Fig. 2c, the number of RD stars in Tycho-2 should be approximately equal to the number of SG stars and is approximately a factor of $3-4$ smaller than the 
total number of giants (RGC-Y+RGC-O+RGB). This conclusion completely agrees with the simulations based on the Besan\c{c}on model of the Galaxy (Robin et al. 2003).
According to these simulations, the number of dwarfs in the sample for $V<10^m$ is approximately equal to the number of subgiants and is a factor of $3-4$ smaller than 
the number of giants; at $V=11^m$, these three categories are represented approximately equally; as $V$ increases, the number of dwarfs exceeds considerably the number of 
remaining stars. It thus follows that our conclusions are valid for Tycho-2 but will unlikely to be valid for 2MASS and other catalogs with fainter stars, although the method 
proposed here is applicable.

We see from Fig. 2c that the number of RGC-Y* stars is much larger than that of RGC-Y and is approximately the same as that of RGC-O. This means that if the Sun is located 
inside the Local Spiral Arm, i.e., in fact, if the distribution density of RGC-Y stars decreases with distance from the Sun, then the ratio of the number of RGC-Y to that of RGC-O and, in
general, the number of young Tycho-2 stars to that of old stars is considerably larger than that in the absence of the Local Arm, which is consistent with the conclusions by 
Girardi et al. (2005).

Figure 2d shows the same distributions as those in Fig. 2c, but all sources of errors in $M'_V$ were taken into account. In this case, the errors in the original
photometry and proper motions and the simulated absence of allowance for the motion to the apex played a major role.

The simulation results in Fig. 2d closely correspond to the Tycho-2 data in Fig. 1d and Gontcharov (2008b). The following qualitative conclusions can be
drawn from Fig. 2d. Using $M'_{V}$
\begin{itemize}
\item does not allow RGC-Y, RGC-O, and RGB stars to be separated but allows a sample of these categories, i.e., a general sample of thin-disk giants
without any admixture of dwarfs, to be produced;
\item allows dwarfs to be selected almost without any admixture of giants;
\item subgiants and thick-disk giants serve as an admixture in selecting both thin-disk giants and dwarfs.
\end{itemize}

Thus, when the reduced proper motions are used, an approximate (with a certain probability) classification of Tycho-2 stars in the range $B-V=0.75^m\div1.25^m$ is possible. 
A star with $M'_V<6^m$, $6^m<M'_V<8^m$, and $M'_V>8^m$ is most likely a giant, a subgiant, and a dwarf, respectively. 
Samples of stars from certain categories with a small admixture of extraneous stars can also be produced.

We see from Fig. 2d that the errors affect noticeably the distribution of stars in $M'_V$ only for SG and RD: the maximum of their distribution in $M'_V$ was
shifted by 0.6$^m$. The distribution of giants in $M'_{V}$ is not affected by the errors. Therefore, the effect of the limitation of the sample of giants in $M'_{V}$ on kinematic
parameters can be successfully simulated without knowing the errors. In other words, despite the use of $M'_{V}$ in selecting stars, we can estimate the biases
of the kinematic characteristics caused by this and study the kinematics of the corresponding sample.

\begin{figure}
\includegraphics{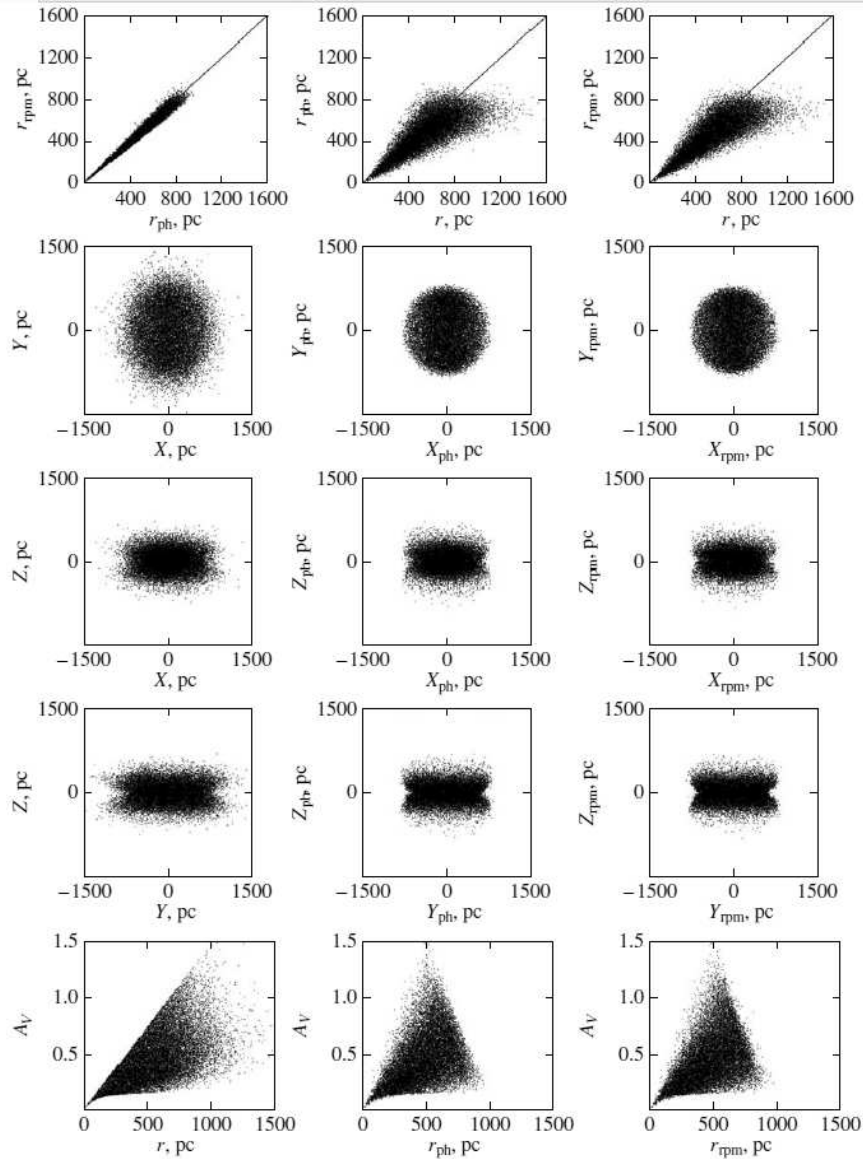}
\caption{Spatial distributions of RGC-Y* stars with the possible effect of the Local Spiral Arm in $r$, $r_{ph}$, and $r_{rpm}$. The extinction
is plotted against $r$, $r_{ph}$ and $r_{rpm}$ at the bottom.
}
\label{rcy}
\end{figure}

\begin{figure}
\includegraphics{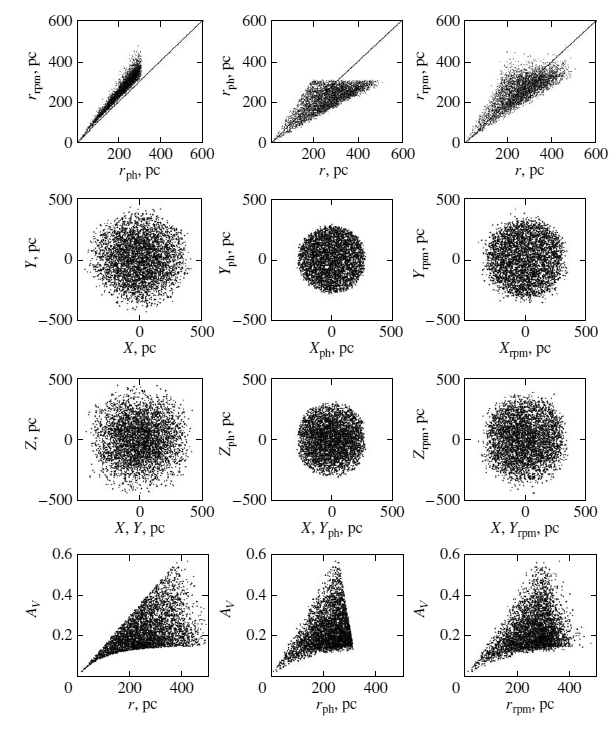}
\caption{Same as Fig. 3 for SG stars. The distributions in $X$ and $Y$ are identical.
}
\label{sg}
\end{figure}

\begin{figure}
\includegraphics{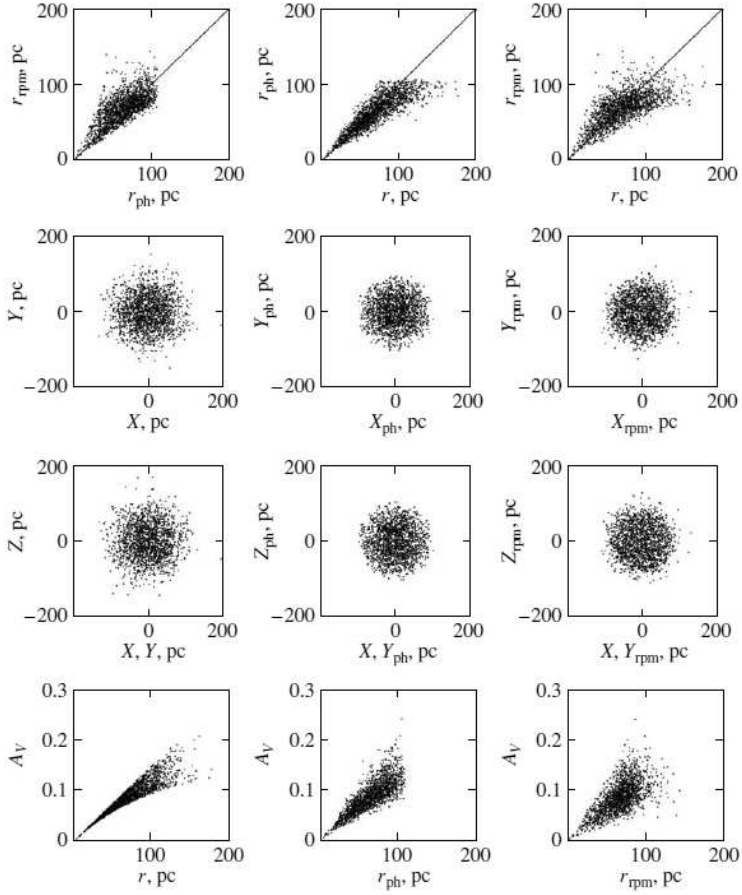}
\caption{Same as Fig. 3 for RD stars. The distributions in $X$ and $Y$ are identical.
}
\label{rd}
\end{figure}

Table 2 presents the mean values and dispersions of $V_l$ and $V_b$ in km s$^{-1}$ for samples of RGC-O stars with different limitations in comparison with the
specified (true) values. The specified mean values differ from zero only due to the Galactic rotation and solar motion to the apex, while all of the remaining
samples are also affected by all of the errors considered here. The second row in Table 2 shows that even in the absence of any selection, except the
selection in apparent magnitude, the characteristics of the sample differ markedly from the true ones, i.e., the observational errors (primarily in the proper
motions) \emph{systematically} affect the kinematic parameters being determined. This unexpected result can be explained by the dependence of the error in the proper
motion on the magnitude specified by Eq. (5) and corresponding well to the reality. Therefore, rejecting the stars with $V>10.6^m$, we reject mostly the stars
with small proper motions, introducing biases into the results.


\begin{table*}
\def\baselinestretch{1}\normalsize\small
\caption[]{Statistical characteristics of the samples of RGC-O stars (affected by the Galactic rotation and solar motion to the apex)}
\label{samples}
\[
\begin{tabular}{lllll}
\hline
\noalign{\smallskip}
 Sample  & $\overline{V_l}$, km/s & $\overline{V_b}$, km/s & $\sigma(V_l)$, km/s  & $\sigma(V_b)$, km/s  \\
\hline
\noalign{\smallskip}
True                                   & 5.7 & 7.5 & 30.0 & 15.0 \\
Without selection in $M'_V$       & 5.8 & 6.2 & 33.6 & 19.0 \\
$(B-V)=0.78\div1.18$, $M'_{V}=3.8\div6.4$ & 5.6 & 7.4 & 32.7 & 20.5 \\
$(B-V)=0.78\div1.18$, $M'_{V}=2.0\div6.4$ & 4.5 & 6.3 & 29.3 & 18.9 \\
$(B-V)=0.78\div1.18$, $M'_{V}=0.0\div6.4$ & 4.2 & 5.8 & 28.4 & 18.2 \\
\hline
\end{tabular}
\]
\end{table*}


The third row in Table 2 presents a sample that almost coincides with the sample from Gontcharov (2008b), where the RGC stars are selected
in the ellipses that fit best into the isoline of equal distribution density of stars on the $(B-V)$ -- $M'_V$ diagram. In this case, the isoline corresponding to the
maximum gradient of the distribution density across it is chosen. This criterion allows RGC to be selected precisely as the stars in the region of an enhanced
distribution density of giants on this diagram. Indeed, comparison of the first and third rows in Table 2 shows that this selection method gives unbiased mean $V_l$
and $V_b$.

The fourth and fifth rows in Table 2 present the results when slow stars of higher luminosities with $2.0<M'_V<3.8$ and $0.0<M'_V<3.8$, respectively,
which, judging by Fig. 2d, are numerous, were added to the sample from Gontcharov (2008b). As expected, the dispersions of $V_l$ and $V_b$ in these cases are closest
to the true ones.

Thus, in certain cases, the biases of the determined kinematic characteristics of a sample limited in color index and reduced proper motion are smaller than
the unavoidable biases caused by the observational errors and the magnitude limitation of the catalog. Consequently, \emph{the samples produced using the reduced
proper motions can be used for kinematic studies}. Irrespective of the sample limitations, the expected biases of the kinematic characteristics being
determined should be derived and taken into account using simulations.

\section*{Photometric and photoastrometric distances}

Whereas the dependences of $M_V$ on $B-V$ and $M'_V$ have been determined correctly, the distances calculated from them will still differ from the true ones
due to the scatter of $M_V$ about the adopted average dependences primarily because of the different ages and chemical compositions of the star within each
category. Figures 3, 4, and 5 show the relations between various quantities for RGC-Y*, SG, and RD (the relations for RGC-Y, RGC-O, RGB, and TDG
are not shown, since they are similar to those for RGC-Y*). Comparison of $r$, $r_{ph}$, and $r_{rpm}$ shows the following: (1) the relative random error in $r_{ph}$ 
and $r_{rpm}$ corresponds to the scatter of $M_V$ about its mean specified in Table 1, for example, 25\% and $\pm0.5^m$, respectively, for RGC-Y*; 
(2) $r_{ph}$ and $r_{rpm}$, on average, agree with $r$ and are even slightly larger than $r$ up to some distance and are systematically lower
further out. This effect is observed for all categories of stars. This is the well-known Malmquist effect considered, for example, by Butkevich et al. (2005).
It arises, because mostly higher-luminosity stars fall into the catalog at large distances outside the region of its completeness. Applying the dependences of $M_V$
on $B-V$ and $M'_V$ found from all stars to them, we obtain $r_{ph}$ and $r_{rpm}$ smaller than the true ones. This also manifests itself in the distance dependence of
extinction: $r_{ph}$ and $r_{rpm}$ for high-extinction stars are appreciably smaller than $r$ -- a kind of a ``horn'' of such stars emerges, which is noticeable when a real sample
of presumed RGC stars from Tycho-2 is analyzed (see Fig. 4 in Gontcharov (2008b)).

In general, the spatial distributions of stars in $r_{ph}$ and $r_{rpm}$ agree well with the specified distributions in $r$ for RGC stars and with the actual ones found in
Gontcharov (2008b). The influence of extinction that removes the stars near the Galactic equator from the catalog is noticeable. Under the effect of the spiral
arm, the set of RGC-Y* stars is elongated along the $Y$ axis and is appreciably more compact in the $XZ$ plane than it is in the $YZ$ plane, which completely
agrees with our analysis of a real sample (see Fig. 5 in Gontcharov (2008b)).

The agreement between the distances $r_{ph}$ and $r_{rpm}$ is good for RGC-Y*, RGC-Y, RGC-O, TDG, and SG and slightly poorer for RGB and RD. This is an
important result, since the photometric and photoastrometric distances have always been used separately, without any comparison with each other, and the
legitimacy of using $r_{rpm}$ has often been called into question. Figure 6 shows the relation between $r_{ph}$ and $r_{rpm}$ for 95852 presumed RGC stars selected from
Tycho-2 in Gontcharov (2008b) (for the remaining stars, the distances inferred from Hipparcos parallaxes were used in Gontcharov (2008b)): we see good
agreement between the distances, as in the simulations under consideration.

The correlation between $r_{ph}$ and $r_{rpm}$ becomes clear after the following transformations. Using Eqs. (2)–(4), the relation
$$ r_{ph}\approx~r_{rpm} $$
can be replaced with
$$ c(B-V)+d\approx~aM'_V+b, $$
where $a$, $b$, $c$, and $d$ are some coefficients. Using Eq. (1), we obtain
$$ cB-cV\approx~aV+5a+5a\log(\mu)-aA_V+b-d. $$
Discarding the constants and neglecting the extinction, we will obtain the proportionality
$$ cB-cV\sim~aV+5a\log(\mu). $$
If $c\approx~-a$, as for RGC-Y, RGC-O, SG, and TDG, then
$$ B\sim-\log(\mu). $$
This proportionality is actually seen in Fig. 7a, which shows the correlation between the $B$ magnitude and $-\log(\mu)$ for the presumed RGC stars selected
from Tycho-2 in Gontcharov (2008b).

If $|c|\gg~|a|$, as for RGB and RD, then the correlation between $B$ and $-\log(\mu)$ is weak. This can be seen from Fig. 7b, which shows $\log(\mu)$ as a function
of $B$ for the presumed RD stars selected from Tycho-2 for $M'_V>11^m$, $\sigma(\mu_{\alpha})<0.003$ arcsec yr$^{-1}$, $\sigma(\mu_{\delta})<0.003$ arcsec yr$^{-1}$, 
$\sigma(B-V)<0.2^m$, $0.7^m<(B-V)<1.4^m$.

Since the dependence of MV on $B-V$ for RGB and RD stars is strong, the photometric distances for these categories are preferred. In contrast, the
photoastrometric distances are preferred for RGC-Y, RGC-O, TDG, and SG with a weak dependence of $M_V$ on $B-V$.

\begin{figure}
\includegraphics{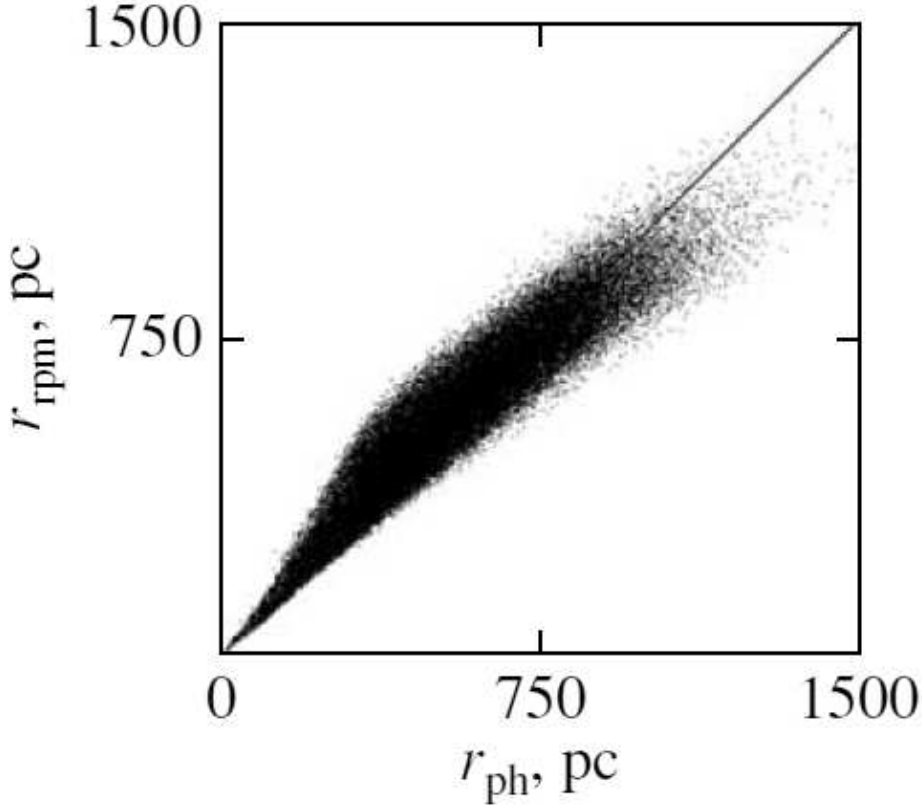}
\caption{Relation between $r_{ph}$ and $r_{rpm}$ for 95852 presumed RGC stars selected from Tycho-2 in Gontcharov (2008b).
}
\label{rrrgc}
\end{figure}

Having considered the general result, let us ascertain how it is affected by various errors. The errors in the photometry, proper motions, and extinction
specified by Eqs. (5), (6), and (7), respectively, turned out to have the greatest effect.

Changing the dispersions of $V_l$ and $V_b$ by $\pm5$ km s$^{-1}$ and the dispersion of the $Z$ distribution of stars by $\pm20$\% of the values listed in Table 1 affects the
results only slightly.

The proper motions used to calculate $M'_V$ can hardly be corrected for the Galactic rotation and solar motion to the apex without any errors, because the
parameters of these motions depend on the sample of stars. However, our simulations show that even if $\mu$ that were partially corrected or completely uncorrected
for the systematic motions are used to calculate $M'_V$ , this makes only slight changes to the results. Only for RD and SG, as for nearby stars, is
the effect of the error in the correction for the motion to the apex noticeable, although it is weaker than the effect of the errors in the photometry and proper
motions. An inaccurate correction for the motion to the apex is one of the reasons why the photometric distances for RD are more accurate than the photoastrometric ones.

The $(B-V)$ -- $M_V$ calibration (4) is, on average, very accurate (but it has a large natural scatter), since it is based not only on the numerous empirical
data but also on the calculations of the stellar evolution theory presented, for example, by Girardi et al. (2000). In contrast, the $M'_V$ -- $M_V$ calibration (3)
cannot be determined theoretically primarily because of the presence of unknown systematic stellar motions in the solar neighborhood and, in general, depends
significantly on the sample of stars. In practice, the coefficients $a$ and $b$ of dependence (3) can be determined only from Hipparcos stars with reliable
trigonometric parallaxes. However, our simulations show that only for RGB are these coefficients significantly different for all RGB stars and the stars closer
than 300 pc. As a result, the distances $r_{rpm}$ for RGB stars, which are anyway systematically smaller than $r$, were reduced additionally. For the remaining categories
of stars, this source of errors is unimportant.

Figure 8 shows the relations between some quantities for RGC-Y*, RGC-O, RGB, TDG, SG, and RD after the introduction of all errors (the relations for
RGC-Y are virtually the same as those for RGC-Y*). Among all of the errors, the errors in the photometry and proper motions play a major role. The errors in the
correction for the motion to the apex and the errors of calibration (3) are also important for SG and RD and RGB, respectively.

\begin{figure}
\includegraphics{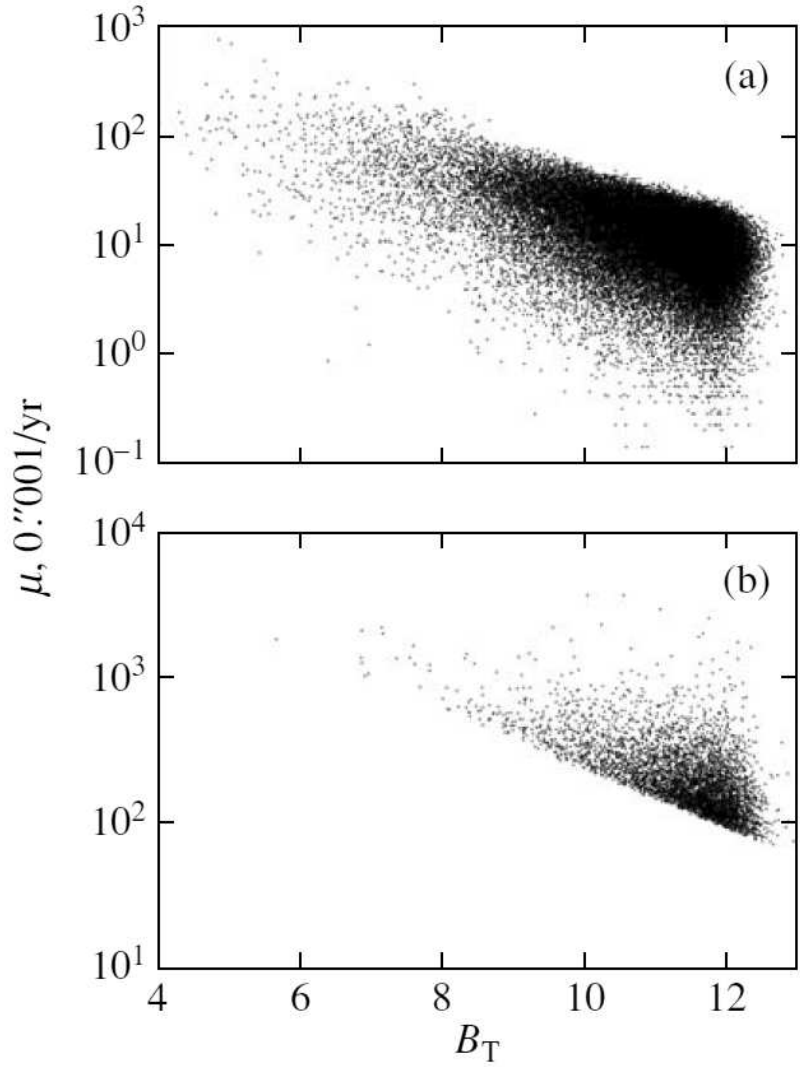}
\caption{Proper motion on a logarithmic scale versus $B_T$ magnitude from Tycho-2: (a) for 97348 presumed RGC stars selected in Gontcharov (2008b) and (b) for 6565
presumed RD stars.
}
\label{bmu}
\end{figure}

\section*{The selection of RGC stars and red dwarfs}

The sample of RGC stars in the optimal (from the viewpoint of minimization of the admixtures) range $0<M'_V<6.4$ will contain $64-71$\% of RGC-Y
and RGC-O, $17-21\%$ of RGB, $9-12\%$ of SG, about 1\% of TDG, and $1-1.5\%$ of RD, depending on the distribution of RGC-Y stars under the effect
of the spiral arm. A rise in the distribution density of stars near the Sun increases the fraction of RGC and reduces the fractions of the remaining categories.

In the selection method applied in Gontcharov (2008b), RGC-Y and RGC-O, RGB, SG, TDG, and RD will account for $60-65$, $18-21$, $13-15$, about 1.5,
and about 2\% of the sample.

Given all errors, the distances $r_{ph}$ and $r_{rpm}$ for the SG stars that are an admixture in the sample of RGC stars agree well with each other but are systematically
larger than the true distance $r$, because the false, as for RGC, dependences (3) and (4) are taken for such SG stars. When the stars are selected by the method
applied in Gontcharov (2008b), the distributions of SG and RGC in $r_{ph}$ and $r_{rpm}$ are approximately the same as those for the stars selected by the condition
$0<M'_V<6.4$.

To reduce the fraction of SG stars in such a sample, it can be limited by some distance: for example, the SG stars will account for only 7\% of the RGC
stars at $r_{ph}<500$ pc (an approximate completeness region of Tycho-2 for RGC) and 9 instead of 12\% in the absence of any limitation at $r_{ph}<700$ pc.
The limitation in $r_{ph}$ is more advantageous than that in $r_{rpm}$. Thus, a sample of RGC-Y, RGC-O, and RGB stars, i.e., thin-disk giants with an admixture
of other stars less than 10\%, can be produced.

We see from Fig. 2d that RD can be selected from Tycho-2. We are planning to make this selection in the next papers. Under the conditions $8<M'_{V}<14$ and 
$(B-V)>0.8^{m}$, no RGC-Y and RGC-O stars will fall into the sample, while RGB, TDG, SG, and RD will account for 0.8, 0.7, 12.4, and more than 86\%, respectively. 
Thus, selecting the stars by the reduced proper motion and the color index, we can produce a sample of red dwarfs with an admixture of other stars less than 14\%.

Given all errors, the distances $r_{ph}$ and $r_{rpm}$ for the SG stars that are an admixture in the RD sample agree well with each other and are systematically
smaller than the true distance $r$. Although RD and SG in such a sample are, on average, at different distances $r$, unfortunately, they are inseparable in $r_{ph}$
and $r_{rpm}$.

A sample of RD stars with a smaller admixture can be obtained if we toughen the selection condition:
$M'_V>9$. RD and SG will then account for 98 and 2\% of the sample, respectively. However, this selection enhances the selection effect in favor of
fast lower-luminosity stars and, as our simulations showed, greatly biases the kinematic characteristics of the sample relative to the true ones.

\begin{figure}
\includegraphics{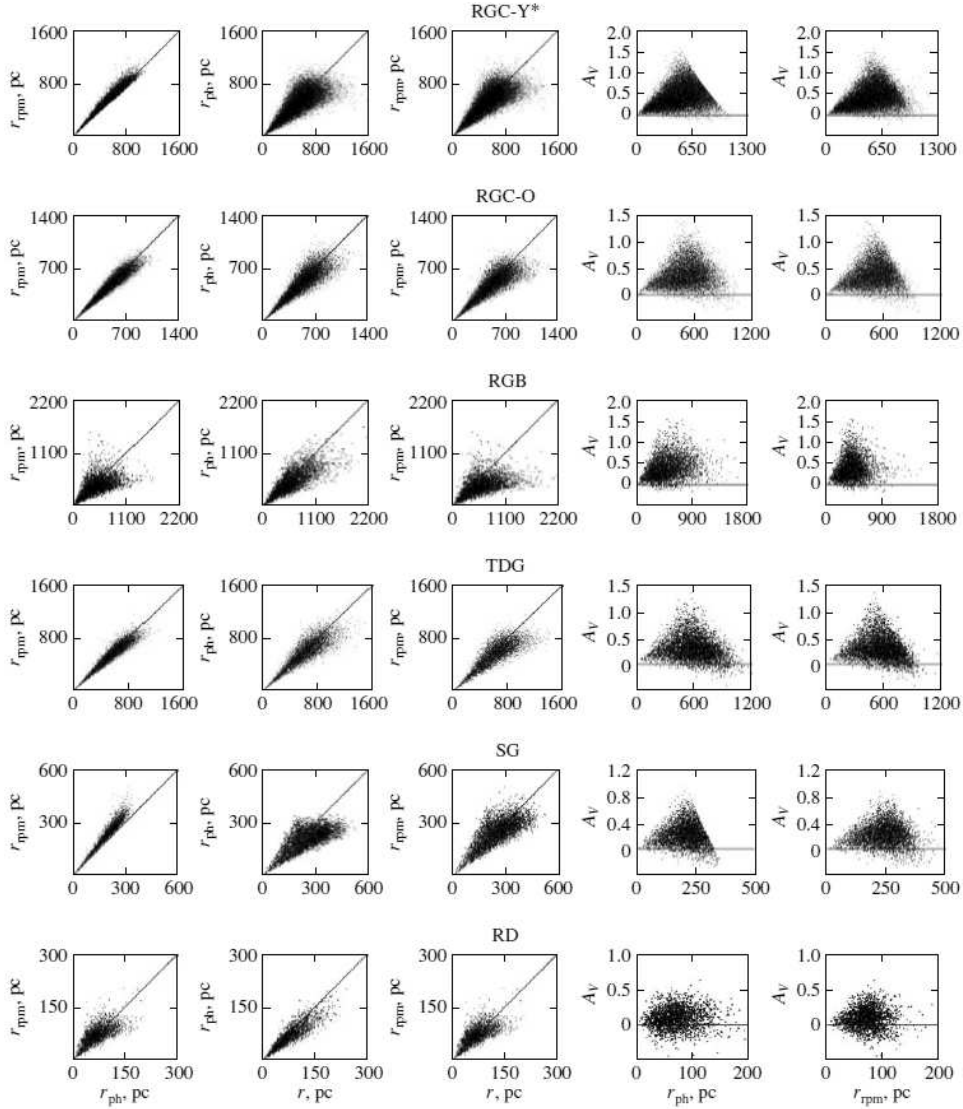}
\caption{Distances and extinctions for various categories of stars under the effect of all the errors considered.
}
\label{err}
\end{figure}

\section*{Conclusions}

Our simulations showed the possibility of an approximate classification and the production of samples of Tycho-2 stars in the range $0.75^m<B-V<1.25^m$ 
based on the reduced proper motion and the color index. In this case, different categories of thin-disk giants cannot be separated, but their total
sample within 500 pc of the Sun with an admixture of other stars less than 10\% can be produced. A sample of red dwarfs within about 100 pc of the
Sun almost without any admixture of other stars can also be produced. An admixture-free sample of Tycho-2 subdwarfs with $B-V<0.75^m$ can probably be produced.

Purity, completeness in some region of space, and symmetry cannot be combined in the same sample without any bias in favor of slow or fast stars.

Our simulations showed that the samples produced using the reduced proper motions could be used for kinematic studies to a no lesser degree than
magnitude-limited samples. In both cases, the kinematic characteristics are biased. Consequently, simulations and allowance for the biases of the quantities
being determined are necessary and possible in both cases.

When simulating the photometric (based on the $(B-V)$ -- $M_V$ calibration) and photoastrometric (based on the reduced proper motion -- $M_V$ calibration) distances,
the main and unexpected result is their correlation. The latter can be explained by the correlation between the apparent magnitude and the logarithm
of proper motion, which is particularly strong for RGC, TDG, and SG and weaker for RGB and RD. As a result, although both distances can be used to
analyze the distributions of stars, the photometric and photoastrometric distances are closer to the true ones, respectively, for RGB and RD stars and for
RGC, TDG, and SG stars. Both distances differ from the true ones primarily because of the natural scatter of absolute magnitudes within each category
of stars, the errors in the original photometry and proper motions, and the Malmquist bias because the catalog is limited in apparent magnitude.

All of the simulation results for RGC stars are in complete agreement with the results of our previous study of a real sample of such stars (Gontcharov 2008b).

All of our qualitative conclusions are valid for any catalog whose characteristics, for example, the photometric and astrometric accuracies and the magnitude
limitation, are close to those of Tycho-2. For catalogs with fainter and, hence, more distant stars, the conclusions can be different, although the presented
method is applicable.

\section*{Acknowledgments}

In this study, I used various results of the Hipparcos Project and data from 2MASS (Two Micron All Sky Survey), which is a joint project of the
Massachusetts University and the IR Data Reduction and Analysis Center of the California Institute of Technology financed by NASA and the National
Science Foundation. I also used resources from the Strasbourg Data Center (France) (http://cdsweb.ustrasbg.fr/). This study was supported by the Russian
Foundation for Basic Research (project no. 08-02-00400) and in part by the ``Origin and Evolution of Stars and Galaxies'' Program of the Presidium of the
Russian Academy of Sciences.


\begin{thebibliography}{99}

\bibitem{belikov} A. N. Belikov, N. V. Kharchenko, A. E. Piskunov, et al., Astron. Astrophys. \textbf{387}, 117 (2002).
\bibitem{butke} A. G. Butkevich, A. V. Berdyugin, and P. Teerikorpi, Mon. Not. R. Astron. Soc. \textbf{362}, 321 (2005).
\bibitem{hip} ESA, Hipparcos and Tycho Catalogues (ESA, 1997).
\bibitem{girardi2000} L. Girardi, A. Bressan, G. Bertelli, et al., Astron. Astrophys. Suppl. Ser. \textbf{141}, 371 (2000).
\bibitem{girardi2005} L. Girardi, M. A. T. Groenewegen, E. Hatziminaoglou, et al., Astron. Astrophys. \textbf{436}, 895 (2005); http://ad.usno.navy.mil/ucac/.
\bibitem{ob} G. A. Gontcharov, Pis'ma Astron. Zh. \textbf{34}, 10 (2008a) [Astron. Lett. \textbf{34}, 7 (2008a)].
\bibitem{rcg} G. A. Gontcharov, Pis'ma Astron. Zh. \textbf{34}, 868 (2008b) [Astron. Lett. \textbf{34}, 785 (2008b)].
\bibitem{tycho2} E. H\o g, C. Fabricius, V. V. Makarov, et al., Astron. Astrophys. \textbf{355}, L27 (2000).
\bibitem{jones} E. M. Jones, Astrophys. J. \textbf{173}, 671 (1972).
\bibitem{parenago} P. P. Parenago, A Course in Stellar Astronomy (GITTL,Moscow, 1954) [in Russian].
\bibitem{robin} A. C. Robin, C. Reyle, S. Derriere, et al. Astron. Astrophys. \textbf{409}, 523 (2003).
\bibitem{rocha} H. J. Rocha-Pinto, S. R. Majewski, M. F. Skrutskie, et al., Astrophys. J. \textbf{615}, 732 (2004).
\bibitem{rybka} S. P. Rybka, Kinemat. Fiz. Neb. Tel \textbf{23}, 102 (2007) [Kinem. Phys. Cel. Bodies \textbf{23}, 70 (2007)].
\bibitem{salpeter} E. E. Salpeter, Astrophys. J. \textbf{121}, 161 (1955).
\bibitem{2mass} M. F. Skrutskie, R. M. Cutri, R. Stiening, et al., Astron. J. \textbf{131}, 1163 (2006); http://www.ipac.caltech.edu/2mass/releases/allsky/.
\bibitem{soubiran} C. Soubiran, O. Bienayme, and A. Siebert, Astron. Astrophys. \textbf{398}, 141 (2003).
\bibitem{veltz} L. Veltz, O. Bienayme, K. C. Freeman, et al., Astron. Astrophys. \textbf{480}, 753 (2008).
\bibitem{wichman} B. A. Wichman and I. D. Hill, Appl. Statist. \textbf{31}, 188 (1982).
\bibitem{ucac} N. Zacharias, S. E. Urban, M. I. Zacharias, et al., Astron. J. \textbf{127}, 3043 (2004).

\end{thebibliography}
\end{document}